\documentclass[twocolumn,floats,floatfix,superscriptaddress,aps,pra]{revtex4}

\usepackage{amsfonts}
\usepackage{amssymb}
\usepackage{amsmath}
\usepackage{calc}
\usepackage{graphicx}
\usepackage{bm}
\usepackage{braket}
\usepackage{verbatim}

\graphicspath{{pics/}}

\newcommand{\titel}{}

\usepackage{hyperref}
\hypersetup{
pdftitle=\titel,
pdfauthor=J. Nehrkorn
pdfsubject={},
pdfproducer={},
colorlinks=false,
pdfborder=0 0 0	
}

\newcommand{\emax}{\epsilon_\mathrm{max}}
\newcommand{\Ham}{\hat{H}}
\newcommand{\Dt}{\Delta t}

\begin{document}

\title{Staying adiabatic with unknown energy gap}

\author{J. Nehrkorn}
\affiliation{Institut f\"{u}r Quanteninformationsverarbeitung,
Universit\"{a}t Ulm, Albert-Einstein-Allee 11, 89081 Ulm, Germany.}
\author{S. Montangero}
\affiliation{Institut f\"{u}r Quanteninformationsverarbeitung,
Universit\"{a}t Ulm, Albert-Einstein-Allee 11, 89081 Ulm, Germany.}
\author{A. Ekert}
\affiliation{Mathematical Institute, University of Oxford, 24-29
  St. Giles’, Oxford, OX1 3LB, United Kingdom}
\affiliation{Centre for Quantum Technologies,
National University of Singapore, 3 Science Drive 2, 117543 Singapore.} 
\author{A. Smerzi}
\affiliation{BEC Center, Dipartimento di Fisica
Università di Trento, Via Sommarive 14,
I-38123 Povo, Italy.}
\author{R. Fazio}
\affiliation{
NEST, Scuola Normale Superiore and Istituto Nanoscienze-CNR, Piazza
dei Cavalieri 7, I-56126 Pisa, Italy.}
\author{T. Calarco}
\affiliation{Institut f\"{u}r Quanteninformationsverarbeitung,
Universit\"{a}t Ulm, Albert-Einstein-Allee 11, 89081 Ulm, Germany.}

\begin{abstract}
We introduce an algorithm to perform an optimal adiabatic evolution 
that operates without an apriori knowledge of the system spectrum. By
probing the system gap locally, the algorithm maximizes the 
evolution speed, thus minimizing the total evolution time. We 
test the algorithm on the Landau-Zener transition and then apply it 
on the quantum adiabatic computation of 3-SAT: The result is
compatible with an exponential speed-up for up to twenty qubits with respect to 
classical algorithms. We finally study a possible algorithm improvement
by combining it with the quantum Zeno effect. 
\end{abstract}

\maketitle


Adiabatic evolution has been used as a standard driving tool for
quantum systems evolutions for decades. The Adiabatic
Theorem guarantees that a quantum system, prepared in the ground state 
of a time-dependent Hamiltonian, will stay close
to its instantaneous ground state if the Hamiltonian governing the
evolution is varied slowly enough, i.e. adiabatically~\cite{qmbook}. 
The most ubiquitous application of this theorem is the solution of 
quantum mechanical time evolutions in terms of simpler static problems: 
if the adiabatic condition is fulfilled during the evolution, 
i.e. the Hamiltonian is varied slowly with respect to the instantaneous
energy gap, one knows that the quantum system is described by the
instantaneous ground state of the Hamiltonian. A more recent
application of the adiabatic theorem has been pointed out  
to define a new quantum computation model, adiabatic quantum 
computation (AQC) or quantum annealing~\cite{Farhi2001, santoro02}. Indeed, if it is possible to define 
a Hamiltonian whose initial ground state is easy to
prepare and whose final ground state encodes the solution to a
computational problem, the system can be used as a quantum
computer, with the same computational power as the circuit model
quantum computation~\cite{aharonovSIAM08}. In the AQC model, 
the computational complexity is defined via the scaling of 
the time to reach the final state (containing the solution to 
the problem), as a function of the size of the input to the problem. 
Even though, there have been discussions about the consistency of the
adiabatic theorem and the sufficiency of its conditions
~\cite{Marzlin2004,Du2008,Amin2009, TongPRL10}, since the introduction
of AQC, different interesting applications have been 
found, like for example the Grover Search problem where the 
the well known quadratic speed-up can be found analytically by locally probing
the system adiabaticity~\cite{mizel07,raoPRA03,pengPRL08,Roland2002}. 

\begin{figure}[tb]
\includegraphics[angle=0,width=.49\textwidth]{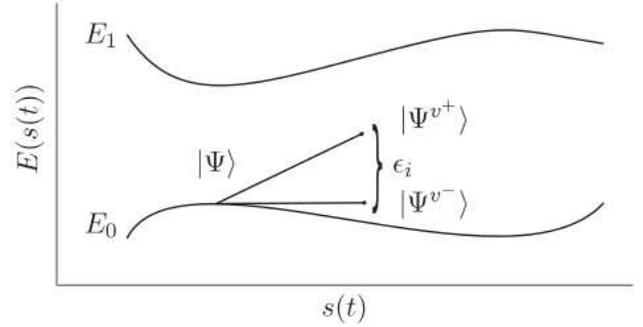}
\caption{Stepwise propagation scheme with two different velocities
  $v_+$ and $v_-$. The state propagated with higher velocity is
  further away from the ground state, the state with smaller velocity
  (i.e. more adiabatic) stays closer to the ground state. After each
  step $i$ the variation $\epsilon_i$ between the two propagated
  states is evaluated and used to update the velocities to remain
  within a predefined error bound such that $\epsilon_i = \emax$.} 
\label{fig:scheme}
\end{figure}
As stated before, the simplest formulation of the adiabatic 
condition states that the time needed to perform an adiabatic 
evolution scales as the inverse of the minimal energy gap between the
ground and the first excited state~\cite{qmbook}. The problem of
finding the time-optimal AQC has been put forward recently and
reformulated also in geometrical 
terms~\cite{RezakhaniPRL09,RezakhaniPRA10,RezakhaniPRA10-1}. However, 
in all these standard approaches it is necessary to know the 
instantaneous spectrum of the system under consideration. 
This task might be very difficult or even impossible, 
like for example when dealing with quantum systems composed by a
large number of particles. Moreover, there are classes of problems,
such as the k-SAT studied here, where different instances of different 
Hamiltonians have to be considered: each of them with a different 
energy spectrum to be computed. 
Very recently, Quan and Zurek introduced a method to
circumvent this problem by using a quenched echo~\cite{quenchecho}. 
The idea is to propagate the ground state of the system forward 
and backward in time comparing the resulting states to evaluate 
the adiabaticity of the evolution. Here, we introduce another 
scheme to find an adiabatic evolution of a given system 
completely independently of any knowledge of the instantaneous 
spectrum: we probe locally (in
time) the adiabaticity of the system and adjust the local speed to
stay close to the instantaneous ground state.  We apply this technique
to the Landau Zener transition~\cite{Zener1932} as testbed and then
to the computationally hard problem of 3-SAT. Finally, as expected
and already shown in a similar setting in~\cite{childsPRA02}, we show that 
combining the algorithm with the quantum Zeno effect~\cite{Misra1977} 
the adiabaticity of the evolution can be improved. 


We consider a quantum system described by a time dependent 
Hamiltonian $\Ham(s(t))$ initially prepared in its ground state.
The Hamiltonian parameter depends on time according to a 
given function $s(t)$ that increases monotonically 
from $s(0)=0$ to $s(T)=1$, where $T$ is the
total evolution time. Our goal is to find the fastest possible way to
perform such transformation while still obtaining an adiabatic transformation 
within a given error $\epsilon$. To find such evolution, without using
any apriori knowledge of the system spectrum and instantaneous ground state, 
we evolve the system state $\ket{\Psi(t)}$ by varying the system
parameter from $s(t)$ at time $t$ with velocity  $v= \Delta s/ \Delta t$,
with two different small velocities $v_\pm = v(1 \pm \delta)$. In general, 
the system ends in two slightly different states that we  
compare to evaluate the adiabaticity of the evolution. Indeed, 
since the energy spectrum is bounded from below, the two states 
are some excited states above the ground state. If the two 
states are almost equal, it means that both velocities $v_\pm$ satisfy the 
adiabaticity condition and the system is in the instantaneous 
ground state, that is, we are performing an adiabatic evolution. 
It is then possible to increase the velocity $v$ to obtain
the fastest possible adiabatic transformation. The limiting case where
the error induced by the transformation with velocity $v_-$
is below the threshold while the other, induced by $v_+$, is above
can be used to determine
the maximal velocity allowed for an adiabatic transformation within a
certain tolerance. 

The idea presented above can be recast in a simple algorithm as
 illustrated in Fig. \ref{fig:scheme}: 
 
\begin{enumerate}\setlength{\itemsep}{1pt}
    \item At $t=0$ prepare the system in the ground state of
      $\Ham(0)$, i.e. $\ket{\Psi(0)} = \ket{\Phi^0(0)}$, 
    \item \label{step:propagate}Propagate the system wave function
      from $\ket{\Psi(t_i)}$ to $\ket{\Psi^{(\pm)}}$, changing $s$
      to $s+\Delta s$ with $v_i^\pm =v_i(1\pm\delta)$ in time
      $\Dt^\pm=\Dt/{(1\pm \delta)}$.
    \item \label{step:evaldelta} Evaluate the distance $\epsilon_i$ between $\ket{\Psi^{(\pm)}}$.
    \item \label{step:updatev} Maximize $v_i$, such that $\epsilon = \epsilon_{\mathrm{max}}$.
    \item Repeat \ref{step:propagate} to \ref{step:updatev} until $s(t) = s(T) = 1$.
\end{enumerate}

The free parameters $\emax$ and $\delta$ are the maximum error that we allow per step,
and the difference between the propagating velocities. By choosing $\Dt$
and $\emax$ appropriately, it is possible to control the final
evolution time $T$ and thus the final fidelity $F(T)=\left|\braket{\Phi^0(T)|\Psi(T)}
\right|^2$. Indeed, the overall final excitation of the evolved state
is a function of the errors accumulated during the evolution. 

The error or distance per step $\epsilon_i$ between the states
$\ket{\Psi^{(\pm)}}$ can be quantified by different figures of
merit. Here we choose the energy difference between the resulting
states 
\begin{equation}
\epsilon_1 = \Delta E = \braket{\Psi^{(+)}|H|\Psi^{(+)}}
  -\braket{\Psi^{(-)}|H|\Psi^{(-)}}
\label{eq:errorperstepE}
\end{equation}
as a natural measure of the adiabaticity of the process. 
We also used the Quantum Fisher Information (QFI) for pure states
~\cite{Braunstein1994, Paris2009}:
\begin{equation}
\epsilon_2 = F_Q^+ -F_Q^- 
\label{eq:qfipure}
\end{equation}
where $F_Q = 4 (\Delta H )^2 = \braket{H^2} - \braket{H}^2$.
The QFI can be seen as a statistical distance between 
quantum states and is a more elaborate method of comparison
than the energy difference. However, as we show in the following, 
the final result is not crucially dependent on the choice 
of the figure of merit. To avoid unphysical ``kinks" 
in the evolution parameter (when the value of the velocity $v$ 
changes discontinuously from $v_i$ to $v_{i+1}$) that would introduce spurious
excitations in the system, we smooth the transition using 
the error function $v(t) = (v_{i-1} - v_i)
\mathrm{erf}[9 (t-t_{i-1})/\Delta t] +
1/2 (v_i+v_{i-1})$ for $t \in [t_{i-1},t_i]$.

\begin{figure}[tb]
\includegraphics[angle=0,width=.49\textwidth]{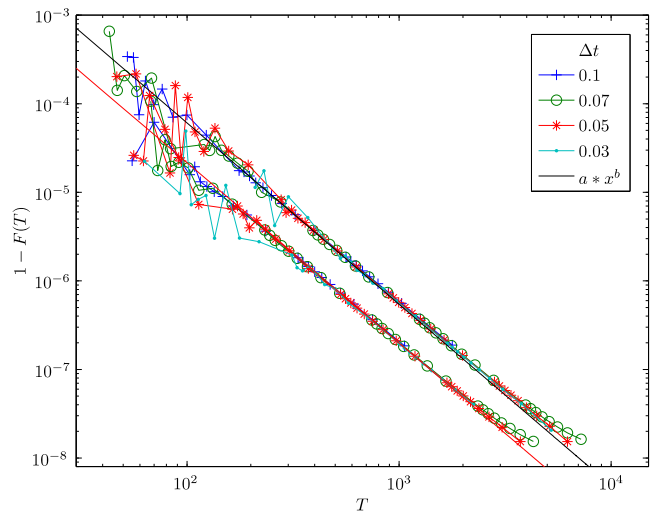}
\caption{Infidelity $1-F(T)$ as a function of the total running
  time $T$ for time step $\Dt=0.1$ (blue crosses), $\Dt=0.07$ (green
  circles), $\Dt=0.05$ (red stars), $\Dt=0.03$ (light blue spots) and
  different errors $\emax= 10^{-10}, \dots, 5 \cdot 10^{-5}$, $\delta=0.8$. 
  Upper (lower) curves are obtained using the energy difference $\Delta E_i$ 
  (Quantum Fisher Information $F_Q$ ) as a figure of merit.
  The black (red) line is a power-law fit $y=a\cdot x^b$
  with $a=0.76$ and $b=-2.05$ ($a=0.26$ and $b=-2.04$).} 
\label{fig:lz}
\end{figure}

First of all, as a testbed of the proposed algorithm,
we search for the optimal adiabatic Landau-Zener (LZ) transition
showing its effectiveness. The adimensional two-level 
time dependent LZ Hamiltonian is given by
\begin{align}
  H_{\mathrm{LZ}}/J  =  \sigma_x + s(t) \sigma_z
\end{align}
where $\sigma_z$ and $\sigma_x$ are the Pauli 
spin matrices~\cite{Zener1932} (we set from now on $\hbar=1$).
We control the adiabatic transformation between the initial and final
ground states of the Hamiltonian $\ket{\Phi^0(0)}$ and
$\ket{\Phi^0(T)}$ by applying the algorithm introduced
above. In Fig.~\ref{fig:lz} we report the final infidelity as 
a function of the total evolution time $T$ for different values 
of the time step $\Dt$ and local error $\emax$, obtained using 
both the energy difference and the Fisher information as a local
measure of error. As it can be clearly seen, in a wide range of parameters 
the data collapse onto two distinct lines (corresponding to the
different local error measure $\epsilon$) following the power law 
\begin{equation}
1-F(T) \sim T^{-2} \sim \frac{\epsilon_{\mathrm{max}}}{\Delta t^2}.
\label{scaling}
\end{equation} 
Fluctuations around this scaling law increase for shorter times,
however the agreement is almost perfect on two orders of magnitude in
time and on four in the infidelity. 
As expected, longer times $T$ allows for better fidelities and
the total evolution time $T$ is determined by the choice of the algorithm
parameters $\epsilon_{\mathrm{max}}$ and $\Delta t$. 
We stress that the results slightly depend on the error measure
$\epsilon_i$ as only the prefactor significantly changes, suggesting 
that the outcome of the algorithm is independent on the details of the
error measure. The important information is whether 
$\ket{\Psi^{(\pm)}}$ are different, not how this is measured.
Similar results have been obtained on different systems, like the
adiabatic version of the Grover search algorithm (data not shown). 
Note that, from the fit of Fig.~\ref{fig:lz} while using the Fisher
information 
($\epsilon_{\mathrm{max}}=F_Q^+$, assuming that the Fisher information
of the state $\ket{\Psi^{-}}$ can be neglected, i.e. that the latter is nearer to
the ground state),  
we obtain that the total error is equal to 
\begin{equation}
1-F(T) \simeq \frac{F_Q}{4 \Delta t^2}.
\label{scaling1}
\end{equation} 
We can compare Eq.(~\ref{scaling1}) with the standard perturbative 
expression for the infidelity of a slightly perturbed state~\cite{Braunstein1994}:
\begin{equation}
1-F(\tau) = \frac{F_Q \tau^2}{4},
\label{scaling2}
\end{equation}
obtaining the relation $\tau= 1/\Delta t$: allowing a maximal error
$\epsilon_{\mathrm{max}}$ every time $\Delta t$ results in an
overall perturbative process with effective time $\tau$. That is, if
$\Delta t \to \infty$ the process is exact.

\begin{figure}[tb]
\includegraphics[angle=0,width=.49\textwidth]{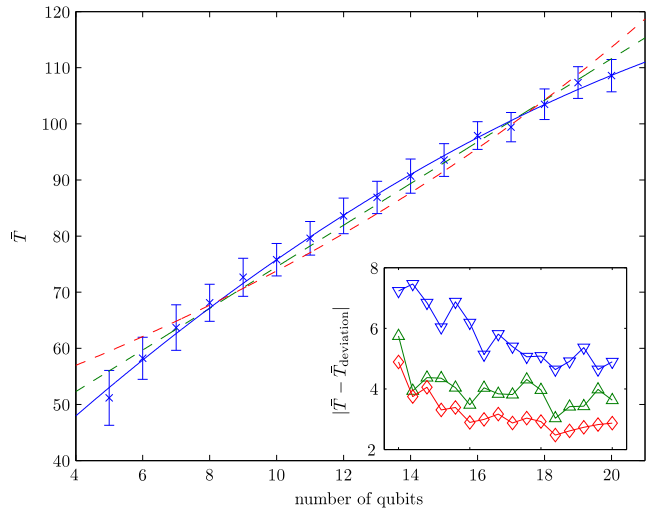}
\caption{Mean total running time $\bar{T}$ for 100 different instances
  of 3-SAT as a function of the  number of qubits $n=5,\dots,20$ and
  $\epsilon_{\mathrm{max}}= 3.4 \cdot 10^{-6}$, $\Delta t =0.1$, $\delta=0.5$. The fits are
  a linear (dashed green), quadratic (solid blue) and an exponential
  function (dashed red). The adjusted coefficient of determination
  $R^2$ are $.98824$, $.9974$ and  $.9627$ for the linear, quadratic
  and exponential respectively. \emph{Inset:} Distance of ten best
  (blue) (ten worst (green)) instances from $\bar{T}$ and
  distribution standard deviation $\sigma$ (red) for $n=4,\dots,20$.} 
\label{fig:sat}
\end{figure}

We then apply the proposed algorithm to an adiabatic quantum
computation used to solve a paradigmatic classically hard problem
--$k$-SAT-- showing how it allows to find a solution with a scaling
with the input size that could be exponentially faster than the
classical one. Given $n$ Boolean variables, the disjunction
of $k$ variables (possibly negated) defines a \emph{clause}: 
the conjunction of $m$ clauses is an \emph{instance} of the 
$k$-SAT problem.
The search for  an assignment for all variables such that all clauses 
of the instance are satisfied at the same time defines the satisfability problem.
The complexity of this problem heavily depends on the ratio of clauses 
and variables $\alpha =m / n$ and for classical algorithms, the
critical value for which the satisfability problem becomes hard is
around $\alpha = 4.3$~\cite{Selman1996}.  
To find the hard instances to test our algorithm, we created sets 
of thousand instances of $3$-SAT for different values of $\alpha$. 
We then solved them with the classical DPLL algorithm
and chose the classically hardest clauses~\cite{Davis1960}. These
clauses showed an exponential increase in the number of 
iterations of the DPLL-algorithm for $\alpha \sim 4.4$ already with
moderate system sizes that can be simulated in the quantum
case~\cite{Davis1960}. The corresponding quantum problem is defined 
along the lines of~\cite{Farhi2001} and simulated via adiabatic 
quantum computation using the algorithm presented before. 
In Fig. \ref{fig:sat} we show the scaling of the total running time $T$
as a function of the number of qubits. Although the scaling is still not 
conclusive, the best fit to the data is a sub-exponential fit, namely
a quadratic function which is in agreement with the results in
~\cite{Farhi2001, Young2008}. The inset shows the difference
between the ten best (ten worst) instances and the mean total running
time $\bar{T}$ and the standard deviation $\sigma$ as a function of
the number of qubits. Note that the width of the distribution for the 
total running time appears to reach a constant value as a function of
the number of qubits: thus, the scaling of the average running time
determines the complexity of the algorithm, giving an indication 
that no pathological clauses exist.

Finally we investigate the possible combination of the presented
algorithm with the Quantum Zeno effect, where a quantum system can 
be frozen in its state by repeated measurements~\cite{Misra1977}. 
It is somehow natural to combine it with our approach, by 
introducing a periodic measurement of the energy of the system at
regular time intervals $t_\mathrm{Z}$. At 
each measurement there is a probability to remain 
in the instantaneous ground state given by $P_i =
\left|\braket{\Psi|\Phi^0(t)}\right|^2$; the 
overall probability to end in the final ground state is the product
$P_{n_\mathrm{z}} = \prod_i^{n_\mathrm{z}} P_i$, where $n_\mathrm{z}=T/t_\mathrm{Z}$
is the number of Zeno measurements. As a manifestation of
the Quantum Zeno effect, both $P_{n_\mathrm{z}}$ and the final fidelity 
should increase for larger
$n_\mathrm{z}$~\cite{childsPRA02}. Moreover, for faster,
i.e. less adiabatic evolutions, the effect should be more significant 
as the projective measurement will ensure that the system remains 
in the ground state. 

\begin{figure}[tb]
\includegraphics[angle=0,width=.49\textwidth]{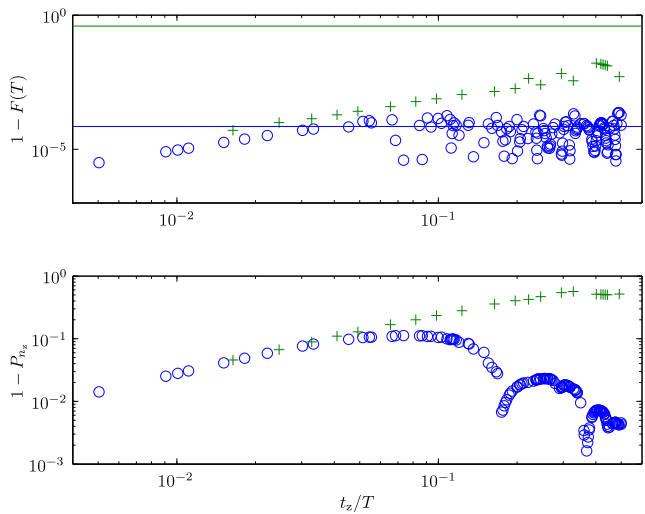}
\caption{\emph{Upper panel:} Final infidelity $1-F(T)$ as a function of the 
  time between Zeno measurements $t_\mathrm{z}/T$ (symbols) normalized
  to the total time $T$ and without Zeno measurements $n_\mathrm{z}=0$
  (lines) for different total time  $T=12.2, \delta t=0.1$ (green
  data) and $T=99.2,  \delta t=0.1$ (blue data) and $\delta=0.8$. 
  \emph{Lower panel:} Final excitation probability 
  $1-P_{n_\mathrm{z}}$ (Probability of algorithm failure)
  as a function of $t_\mathrm{z}/T$ for  $T=12.2$ (green corsses) and 
  $T=99.2$ (blue circles).} 
\label{fig:zeno}
\end{figure}

In Fig.~\ref{fig:zeno} the infidelity $1-F(T)$ and probability
$1-P_{n_\mathrm{z}}$ to leave the ground state are plotted against
the normalized time between Zeno measurement $t_\mathrm{z}/T$ 
for two different total evolution times $T$. The results clearly
indicates that the two total times chosen, almost one order of magnitude
apart, corresponds to a non-adiabatic and an adiabatic evolution.
We normalize the Zeno time $t_\mathrm{z}$ to fairly compare 
such different time scales. The upper panel of Fig.~\ref{fig:zeno} shows that the 
final infidelity decreases for shorter intervals $t_\mathrm{z}/T$, 
i.e. more measurements, as one would expect. However, when compared 
with the final infidelity without Zeno measurements (full lines) 
a big improvement can be seen in the region $0.01 \lesssim
t_\mathrm{z}/T  \lesssim 0.1$ for the 
non-adiabatic case (more than four orders of magnitude) while 
in the other case about one order of magnitude can be gained. 
In the lower panel of Fig.~\ref{fig:zeno} we report the
corresponding probability of the algorithm failure
$1-P_{n_\mathrm{z}}$ as a function of  $t_\mathrm{z}/T$. Here, the two
different kinds of evolutions studied are clearly visible: in the case
of the adiabatic evolution, the probability of exciting the system is
greater in presence of the Zeno measurements, while in the
non-adiabatic case the Zeno measurements helps the systems to remain
int he ground state. Indeed, in the latter case, for  $0.01 \sim
t_\mathrm{z}/T$ there is an improvement of the final fidelity of about
four orders of magnitude with a chance of failure of only a few
percent: for very fast evolutions exploiting  
Zeno effect is a successful strategy to improve the algorithm convergence. 

In conclusion, we introduced an algorithm to optimize an adiabatic
transition for an arbitrary time dependent quantum system with total
ignorance of the instantaneous eigenstates and the resulting gap of
the system. Numerical simulations showed an expected correlation between
the total running time and the final fidelity which does not largely
depend on the comparison method used for the evaluation of the
adiabaticity of the evolution of each step. We applied the 
algorithm to the 3-SAT problem and reproduced the results 
obtained by~\cite{Farhi2001}, suggesting an exponential speed-up of
the adiabatic computation with respect to the classical one. Finally, 
we combined the algorithm with the quantum Zeno effect to further
improved the results for very fast (non-adiabatic) transitions.  
As a final note, we mention that the algorithm effectively 
measures the size of the instantaneous gap, thus it can be used 
to investigate systems with a closing gap, also in combination with
numerical techniques such as tensor network methods where the full
knowledge of the system spectrum is out the computational capability
of actual classical computers. 


We acknowledge support from the EU projects AQUTE, 
the SFB/TRR21 and the BWgrid for computational resources. 


\end{document}